\documentstyle{mn}

\def\etal{{\it et al. }}
\def\deg{\ifmmode^\circ\else$^\circ$\fi}
\def\ltsima{$\; \buildrel < \over \sim \;$}
\def\simlt{\lower.5ex\hbox{\ltsima}}
\def\gtsima{$\; \buildrel > \over \sim \;$}
\def\simgt{\lower.5ex\hbox{\gtsima}}

\newcounter{parentequation}

\begin{document}

\title[Equation of state from Supernovae and CMB Observations]{
Constraining the equation of state of the Universe
from  Distant Type Ia Supernovae and Cosmic Microwave Background
Anisotropies}

\author[G. Efstathiou]{G. Efstathiou \\
Institute of Astronomy, Madingley Road, Cambridge, CB3 OHA.}

\maketitle

\begin{abstract}

The magnitude-redshift relation for Type Ia supernovae is beginning to
provide strong constraints on the cosmic densities contributed by
matter, $\Omega_m$, and a cosmological constant, $\Omega_\Lambda$,
though the results are highly degenerate in the
$\Omega_m$--$\Omega_\Lambda$ plane.  Here we estimate the constraints
that can be placed on a cosmological constant or quintessence-like
component by extending supernovae samples to high redshift. Such
measurements, when combined with constraints from anisotropies in the
cosmic microwave background, could provide an important consistency
check of systematic errors in the supernovae data. A large campaign of
high-z supernovae observations with 10 metre class telescopes could
constraint $\Omega_m$ to a ($1 \sigma$) accuracy of $0.06$ and
$\Omega_\Lambda$ to $0.15$.  A sample of supernovae at redshift $z
\sim 3$, as might be achievable with a Next Generation Space
Telescope, could constrain $\Omega_m$ to an accuracy of about $0.02$
independently of the value of $\Omega_\Lambda$.  The constraints on a
more general equation of state, $w_Q = p/\rho$, converge slowly as the
redshift of the supernovae data is increased. The most promising way
of setting accurate constraints on $w_Q$ is by combining high-z
supernovae and CMB measurements. With feasible measurements it should
be possible to constrain $w_Q$ to a precision of about $0.06$, if the
Universe is assumed to be spatially flat.  We use the recent
supernovae sample of Perlmutter \etal and observations of the CMB
anisotropies to constraint the equation of state in quintessence-like
models via a likelihood analysis.  The $2 \sigma$ upper limits are
$w_Q < - 0.6$ if the Universe is assumed to be spatially flat, and
$w_Q < - 0.4$ for universes of arbitrary spatial curvature. The upper
limit derived for a spatially flat Universe is close to the lower
limit ($w_Q \approx - 0.7$) allowed for simple potentials,
implying that additional fine tuning may be required to construct a
viable quintessence model.

\vskip 0.2 truein
\end{abstract}


\section{Introduction}\label{sec:intro}

The possible discovery of an accelerating Universe from observations
of Type Ia supernovae (Perlmutter \etal 1998, Riess \etal 1999) has
led to a resurgence of interest in the possibility that the Universe
is dominated by a cosmological constant (for a recent review see
Turner 1999). A number of authors have shown how observations of
distant Type Ia supernovae (SN) can be combined with observations of
CMB anistropies to constrain the cosmological constant and matter
density of the Universe (White 1998, Tegmark \etal 1998, Lineweaver
1998, Garnavich \etal, 1998, Efstathiou and Bond 1999, Tegmark 1999,
Efstathiou \etal 1999).  For example, Efstathiou \etal (1999,
hereafter E99) combine the large SN sample of the Supernova Cosmology
Project (Perlmutter \etal 1998, hereafter P98; we will refer to these
supernovae as the SCP sample) with a compilation of CMB anisotropy
measurements and find $\Omega_m = 0.25^{+0.18}_{-0.12}$ and
$\Omega_\Lambda = 0.63^{+0.17}_{-0.23}$ ($95 \%$ confidence errors)
for the cosmic densities contributed by matter and a cosmological
constant respectively. These results are consistent with a number of
other measurements, including dynamical measurements of $\Omega_m$,
the large-scale clustering of galaxies and the abundances of rich
clusters of galaxies (Turner 1999, Bridle \etal 1999, Wang \etal
1999).

\begin{figure*}

\vskip 3.0 truein

\includegraphics{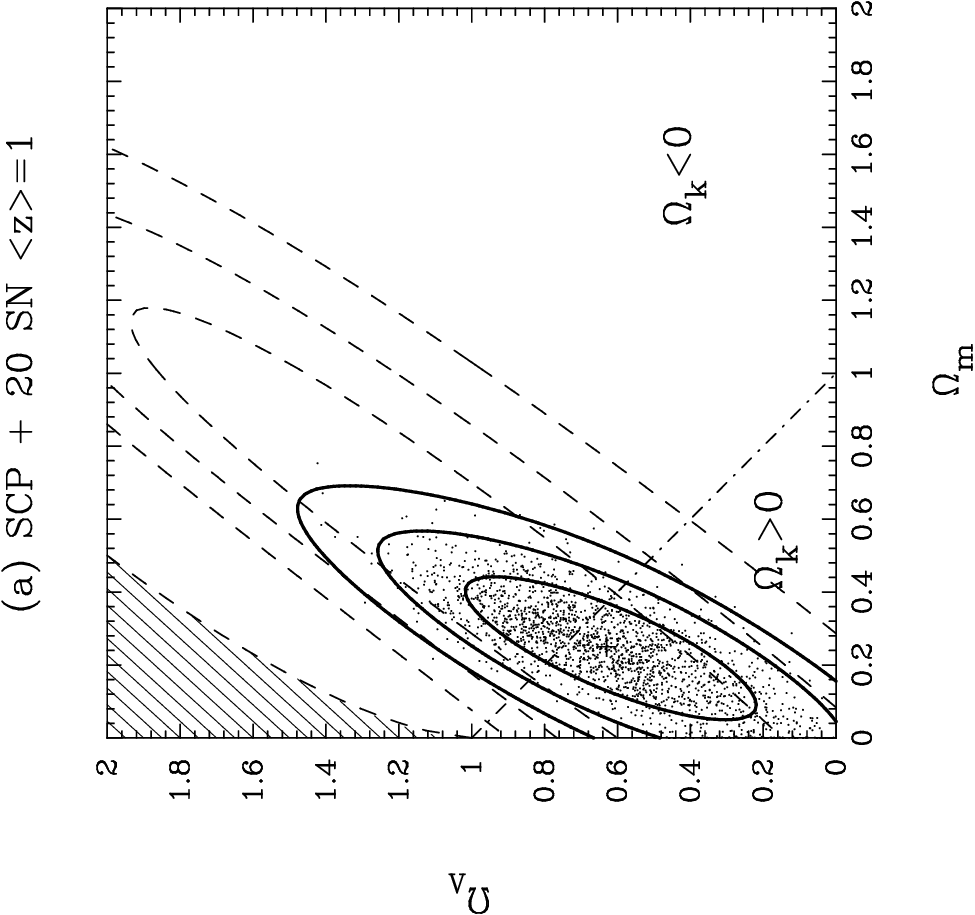}
\includegraphics{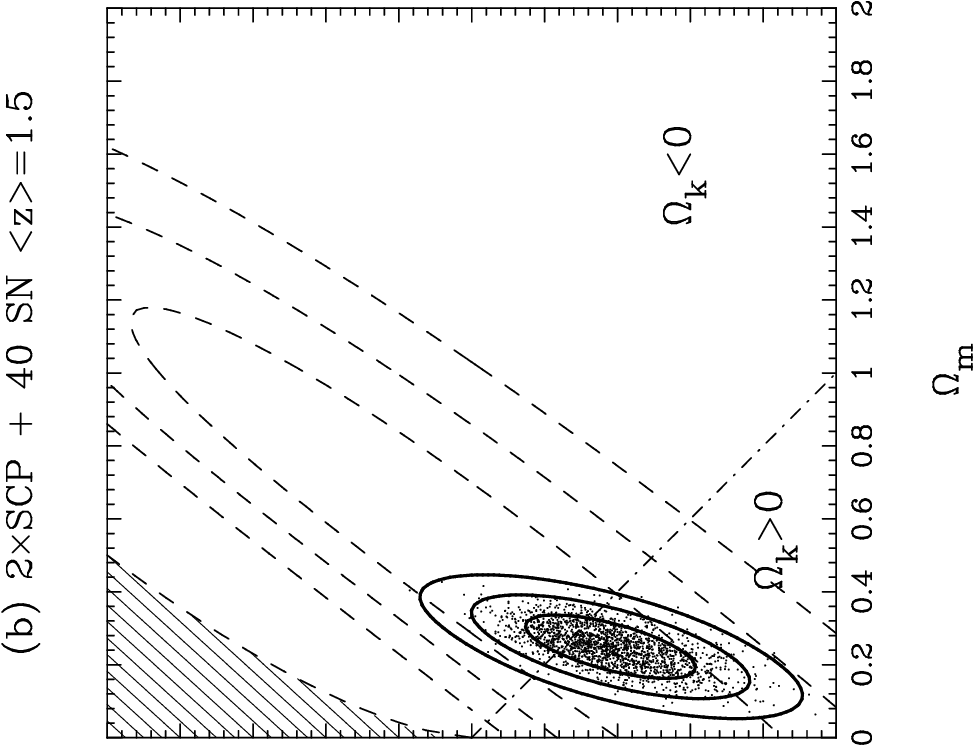}

\caption
{The dashed lines in each panel show $1$, $2$ and $3 \sigma$
likelihood contours in the $\Omega_\Lambda$--$\Omega_m$ plane for the
SCP distant supernova sample as analysed by E99. The solid contours
are derived from the Fisher matrix (equation 4) for the SCP sample
supplemented by 20 SN with a mean redshift of $\langle z \rangle =1$
(Figure 1a) and for twice the SCP sample and $40$ SN with $\langle z
\rangle =1.5$ (Figure 1b). The points show maximum likelihood values
of $\Omega_\Lambda$ and $\Omega_m$ for Monte-Carlo realizations of
these samples, as described in the text.}
\label{figure1}
\end{figure*}

The observational evidence for an accelerating Universe has stimulated
interest in more general models containing a component with an
arbitrary equation of state, $p/\rho = w_Q$ with $w_Q \ge
-1$. Examples include a dynamically evolving scalar field (see {\it
e.g.} Ratra and Peebles 1988 and Caldwell, Dave and Steinhardt 1998,
who have dubbed such a component `quintessence'; we will refer to this
as a `Q' component hereafter) and a frustrated network of topological
defect (Spergel and Pen 1997, Bucher and Spergel 1999). In particular,
Steinhardt , Wang and Zlatev, 1998, have pointed out that for a wide
class of potentials, the evolution of a Q-like scalar field follows
`tracking solutions' , in which the late time evolution is almost
independent of initial conditions.

The purpose of this paper is three-fold. Firstly, to illustrate how
the constraints on $\Omega_m$ and $\Omega_\Lambda$ can be improved by
extending the redshift range of the supernovae samples.  At low
redshifts, the magnitude-redshift relation is degenerate for models
with the same value of the decellaration parameter
$q_0$ ($\equiv {1 \over 2}(\Omega_m - 2\Omega_\Lambda)$). This degeneracy
can be broken by observing supernovae at redshifts $\simgt 1$ (see,
for example, Goobar and Perlmutter, 1995).  Thus, by extending the
redshift range of the current supernovae samples it should be possible
to set tighter limits on $\Omega_m$ and $\Omega_\Lambda$
independently. This is important because there are significant worries
that the SN data may be affected by grey extinction, evolution, or some
other systematic effect.  The consistency of SN constraints on
$\Omega_m$ and $\Omega_\Lambda$ with those derived from the CMB
anisotropy measurements would provide an important consistency check
of systematic errors in the SN data and the interpretation of the CMB
data. Secondly, we estimate the accuracy with which a more general
Q-like equation of state can be constrained by high redshift SN and
CMB data. Thirdly, we use the current SN and CMB anisotropy data to
constrain Q-like models in a spatially flat universe and in a universe with
arbitrary spatial curvature.

\section{Analysis of Models with a Cosmological Constant}

\subsection{Constraints from supernovae at $z \sim 1$}

The predicted peak magnitude-redshift relation is given by
\begin{equation}
m^{\rm pred} (z) =  {\cal M} + 5{\rm log}{\cal D}_L(z, \Omega_m, 
\Omega_\Lambda),
\end{equation}
where ${\cal M}$ is related to the peak absolute
magnitude by ${\cal M} = M - 5 {\rm log}H_0 + 25$.
and ${\cal D}_L = d_L + 5 {\rm log}H_0$
is the Hubble constant-free luminosity distance.
To compute the luminosity distance, we ignore
gravitational lensing and use the standard expression for
a Universe with uniform density (see {\it e.g.}  Peebles 1993),
$$
d_L(z, \Omega_m, \Omega_\Lambda)  =  {c \over H_0} {(1+z) \over \vert
\Omega_k \vert^{1/2}} 
{\rm sin}_k \left [ \vert \Omega_k \vert^{1/2} x(z, \Omega_m,
\Omega_\Lambda) \right ],
$$
$$
x(z, \Omega_m, \Omega_\Lambda) =  
\int_0^z {dz^\prime 
\over 
[\Omega_m (1 + z^\prime)^3 + \Omega_k ( 1 + z^\prime)^2 +
\Omega_\Lambda]^{1/2} }
\quad  (2) 
$$
where $\Omega_k = 1 - \Omega_m - \Omega_\Lambda$ and ${\rm sin}_k =
{\rm sinh}$ if $\Omega_k > 0$ and ${\rm sin_k} = {\rm sin}$ for
$\Omega_k < 0$.  

\begin{table}
\label{tab1}
\centerline{\bf Table 1: Fisher Matrix Errors, $\Omega_m$ and $\Omega_\Lambda$.}
\begin{center}
\begin{tabular}{|cccccc|} \hline
\multicolumn{6}{c} {Supernovae Alone} \\
              &  SCP   & \multicolumn{2}{c}{SCP + 20 SN} &
\multicolumn{2}{c}{2$\times$SCP + 40SN} \\
$<z>$                &          &  $1.0$ & 1.5 & 1.0 & 1.5 \\
$\delta \Omega_m$    &    $0.53$  & $0.130$ & $0.081$ & $0.092$ & $0.057$ \\ 
$\delta \Omega_\Lambda$&  $0.71$  & $0.265$ & $0.218$ & $0.19$ & $0.154$\\
$\delta {\cal M}$   &     $0.056$ & $0.053$ & $0.049$ & $0.035$ & $0.035$\\\hline
\multicolumn{6}{c} {Supernovae + CMB} \\
              &  SCP   & \multicolumn{2}{c}{SCP + 20 SN} &
\multicolumn{2}{c}{2$\times$P98 + 40SN} \\
$<z>$                &          &  $1.0$ & 1.5 & 1.0 & 1.5 \\
$\delta \Omega_m$    &    $0.073$  & $0.055$ & $0.047$ & $0.039$ & $0.033$ \\ 
$\delta \Omega_\Lambda$&  $0.080$  & $0.060$ & $0.051$ & $0.042$ & $0.036$\\
$\delta {\cal M}$   &     $0.046$ & $0.042$ & $0.039$ & $0.030$ & $0.028$\\\hline
\end{tabular} 
\end{center} 
\end{table}

We assume that we observe $N$ supernovae, with peak magnitude $m_i$,
(corrected for k-term, decline rate-luminosity relation, reddening {\it
etc}),  magnitude error $\sigma_i$ and redshift $z_i$, from which we
want to determine a set of parameters $s_k$ by maximising the likelihood
function,
\addtocounter{equation}{1}
\begin{equation}
{\cal L} = \prod_{i=1}^N {1 \over \sqrt{(2 \pi \sigma_i)}} 
{\rm exp} \left
( - {(m_i - m_i^{\rm pred})^2 \over 2 \sigma_i^2} \right ).
\end{equation}
In this section we assume that the parameters $s_k$ are $\Omega_m$, 
 $\Omega_\Lambda$ and ${\cal M}$
(defined in equation 1). An estimate of the covariance matrix,
$C_{ij}$, for these parameters for a given SN data set is given by the
inverse of the Fisher matrix
\begin{equation}
 F_{ij} = \sum_k {1 \over \sigma_k^2} {\partial {m_k^{\rm pred}} \over \partial s_i} 
{\partial {m_k^{\rm pred}} \over \partial s_j}
\end{equation}
(Kendall and Stewart 1979).  The marginalized
error on each parameter (given by $\sqrt C_{ii}$) is listed in Table 1
for several assumed supernova datasets. The column labelled SCP gives
the Fisher matrix errors on $\Omega_m$, $\Omega_\Lambda$ and ${\cal
M}$ derived for sample C ($56$ supernovae) of P98, {\it i.e.}
assuming the magnitude errors, intrinsic magnitude scatter and
redshift distribution of the real sample. The next two columns give
the expected errors for the SCP sample supplemented by 20 supernovae
with a peak magnitude error of $\Delta m = 0.25$ magnitudes and a
Gaussian redshift distribution of dispersion $\Delta z=0.5$ and mean
redshift $\langle z \rangle = 1$ and $1.5$. The upper redshift limit
is close to the maximum for feasible spectroscopic measurements with
$10$ metre-class telescopes (see Goobar and Perlmutter 1995). As these
authors comment, ground based spectroscopy at optical wavelengths
becomes prohibitively expensive for supernovae at higher redshifts
because of the strong K-correction.  The last two columns give the
errors for a sample twice as large as the SCP sample supplemented by
40 supernovae with mean redshift of $1.0$ and $1.5$. We adopt a
background cosmology with $\Omega_\Lambda = 0.63$ and $\Omega_m =
0.25$ as indicated by the joint likelihood analysis of the SCP sample
and CMB anisotropies described in E99.

From Table 1 we see that the Fisher matrix analysis of the SCP sample
gives relatively large errors on $\Omega_\Lambda$ and $\Omega_m$, in
agreement with the likelihood analysis presented by P98. However, by
adding $20$ SN at $z \sim 1$, the errors on $\Omega_\Lambda$ and in
particularly $\Omega_m$ are reduced significantly. The last column
shows that an enhanced SCP sample together with $40$ SN at $z \sim
1.5$ (a formidable, but feasible observing programme) can provide a
tight constraint on $\Omega_m$. The parameters $\Omega_\Lambda$ and
$\Omega_m$ are, of course, highly correlated. This is illustrated in
Figure 1 which shows $1$, $2$ and $3\sigma$ error ellipses in the
$\Omega_\Lambda$-$\Omega_m$ plane after marginalizing over $s_3 =
{\cal M}$ assuming a uniform prior distribution.  (The components of
the new Fisher matrix after marginalization are given by
$F^\prime_{11} = F_{11} -F^2_{13}/F_{33}$, $F^\prime_{22} = F_{22}
-F^2_{23}/F_{33}$, $F^\prime_{12} = F_{12} -F_{13}F_{23}/F_{33}$.) The
points in the Figure show the results of Monte-Carlo calculations,
where we have simulated the observational samples and determined the
parameters $s_i$ by maximising the likelihood function (equation 2).
By diagonalizing the matrix $F^\prime$ we can find the orthogonal
linear combinations $\Omega_\| = a \Omega_m + b \Omega_\Lambda$ and
$\Omega_\bot = b \Omega_m - a \Omega_m$ defining the major and minor
axes of the likelihood contours shown in Figure 1. The distributions
in these orthogonal directions are shown in Figure 2 and compared with
the distributions determined from the Monte-Carlo simulations. The
Monte-Carlo distributions are very close to Gaussians and show that
the Fisher matrix gives an extremely accurate description of the
errors in the $\Omega_m$--$\Omega_\Lambda$ plane.

\begin{figure}

\vskip 5.4 truein

\includegraphics{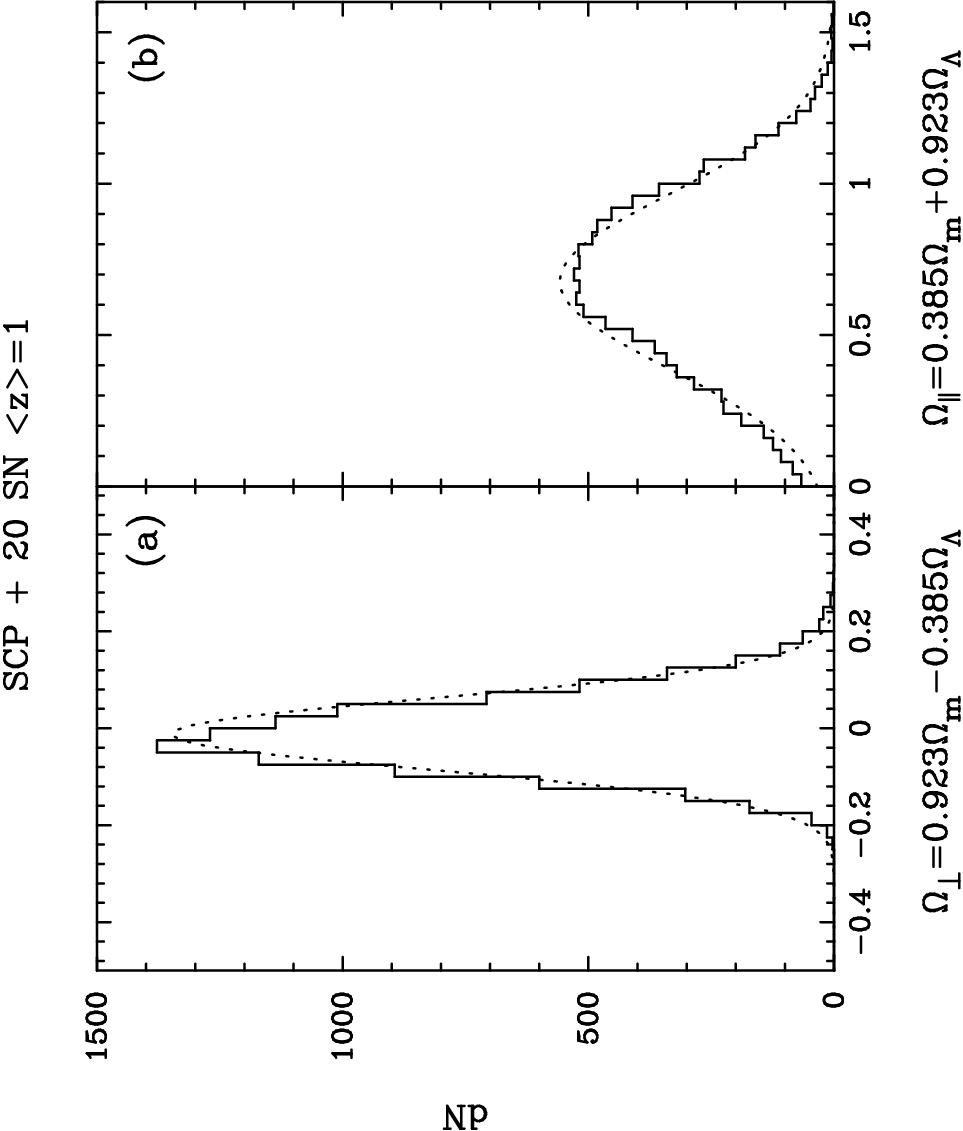}
\includegraphics{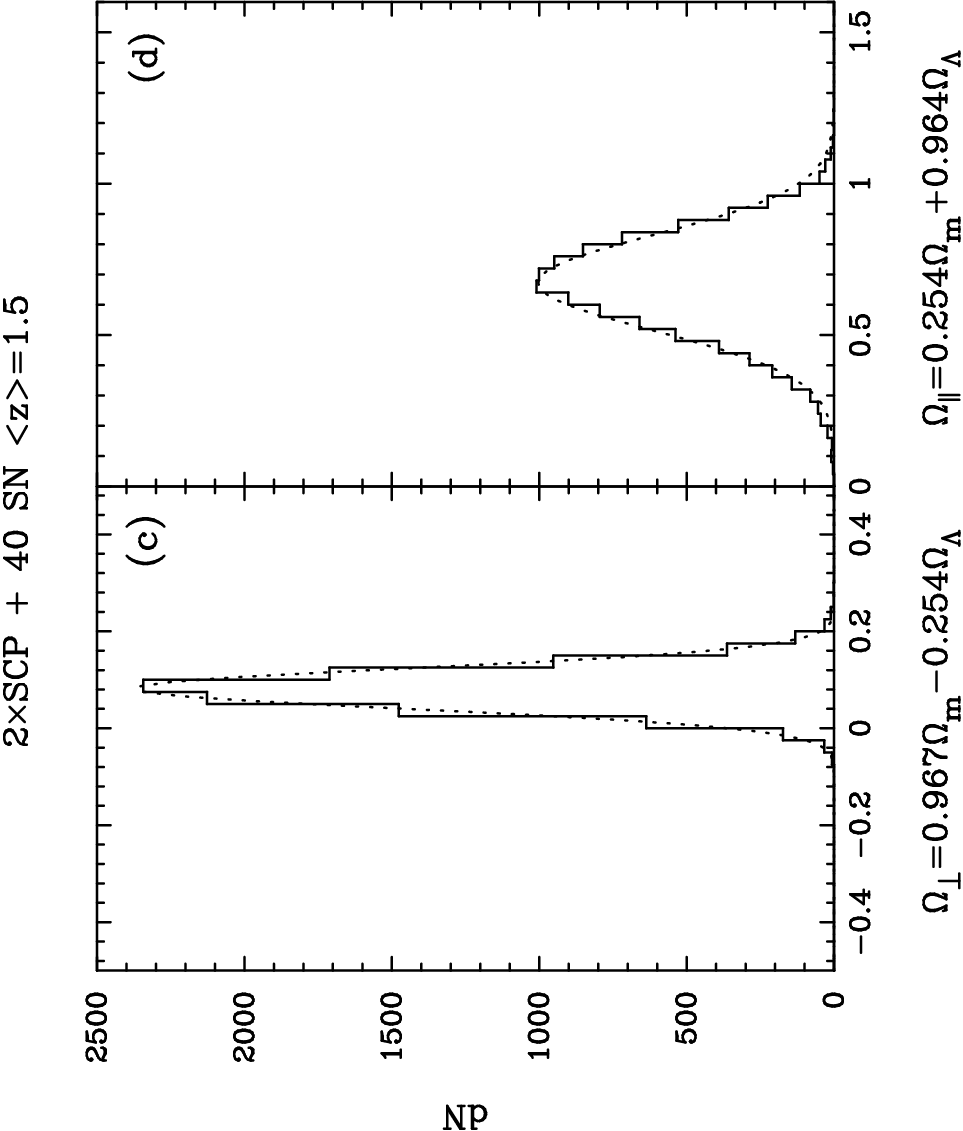}

\caption
{Distributions along the major and minor axes of the likelihood
contours shown in Figure 1. The histograms show the distributions derived
from the Monte-Carlo simulations and the dotted lines show Gaussian distributions with variances determined from the Fisher matrix after marginalizing over
the parameter ${\cal M}$.}

\label{figure2}
\end{figure}

Although the errors in $\Omega_m$ and $\Omega_\Lambda$ are
significantly reduced by the addition of high redshift supernovae over
those of the SCP sample, they are still quite large in the parallel
direction $\Omega_\|$. This means that it is difficult to set tight
limits on $\Omega_\Lambda$ from SN measurements alone. The constraints
on the spatial curvature $\Omega_k$ are even weaker. For example, for
the larger sample shown in Figures 1 and 2, the $1\sigma$ error on
$\Omega_k$ is $\delta \Omega_k = 0.19$. This can be reduced by
extending the range to even higher redshifts (see Section 2.2)
 or by combining the SN data with cosmic
microwave background anisotropies, as has been done by several authors
(White 1998, Lineweaver 1998,  Garnavich \etal 1998, Tegmark 1999, 
E99).

 CMB anisotropy measurements, especially with future satellites such
as MAP and Planck, are capable of setting tight constraints on the
locations of the acoustic peaks in the CMB power spectrum. Following
E99, we define an acoustic peak location parameter $\gamma_D(\Omega_m,
\Omega_\Lambda)$ to be the ratio of the peak position in a model with
arbitrary cosmology compared to that in a spatially flat model with
zero cosmological constant. (This parameter depends weakly on the
matter content of the Universe and on the spectral index of the
fluctuations, but we ignore these small dependences in what follows).
CMB measurements are therefore capable of fixing $\gamma_D$, defining
a degeneracy direction in the $\Omega_\Lambda$--$\Omega_m$ plane given
by
\begin{equation}
\Delta \Omega_\Lambda =  - {(\partial \gamma_D/\partial \Omega_m)
\over (\partial \gamma_D/\partial \Omega_\Lambda)} \Delta \Omega_m,
\end{equation}
(see Efstathiou and Bond 1998). The results in the lower panel of
Table 1 show the Fisher matrix analysis of the SN samples including
the constraint imposed by equation (5). As is well known,
the combination of SN and CMB measurements can break the degeneracy
between $\Omega_\Lambda$ and $\Omega_m$ and it should be possible to
determine these parameters with an error of less than $0.04$ with an
enlarged supernova sample assuming, of course, that systematic errors
are unimportant. 

Although the errors on $\Omega_\Lambda$ from SN measurements alone
converge relatively slowly as the redshift range is increased,
consistency of the cosmological parameter estimates provides a strong
test of systematic errors in the SN data. If we believe that
systematic errors are unimportant, and that our interpretation of the
CMB anisotropies (in terms of adiabatic CDM-like models) are correct,
then current data already constrain $\Omega_m$ and $\Omega_\Lambda$ to
high precision (see Fig 5 of E99). Consistency requires that the
likelihood contours for a high redshift supernova sample converge to
the same answer.

\begin{figure}

\vskip 3.0 truein

\includegraphics{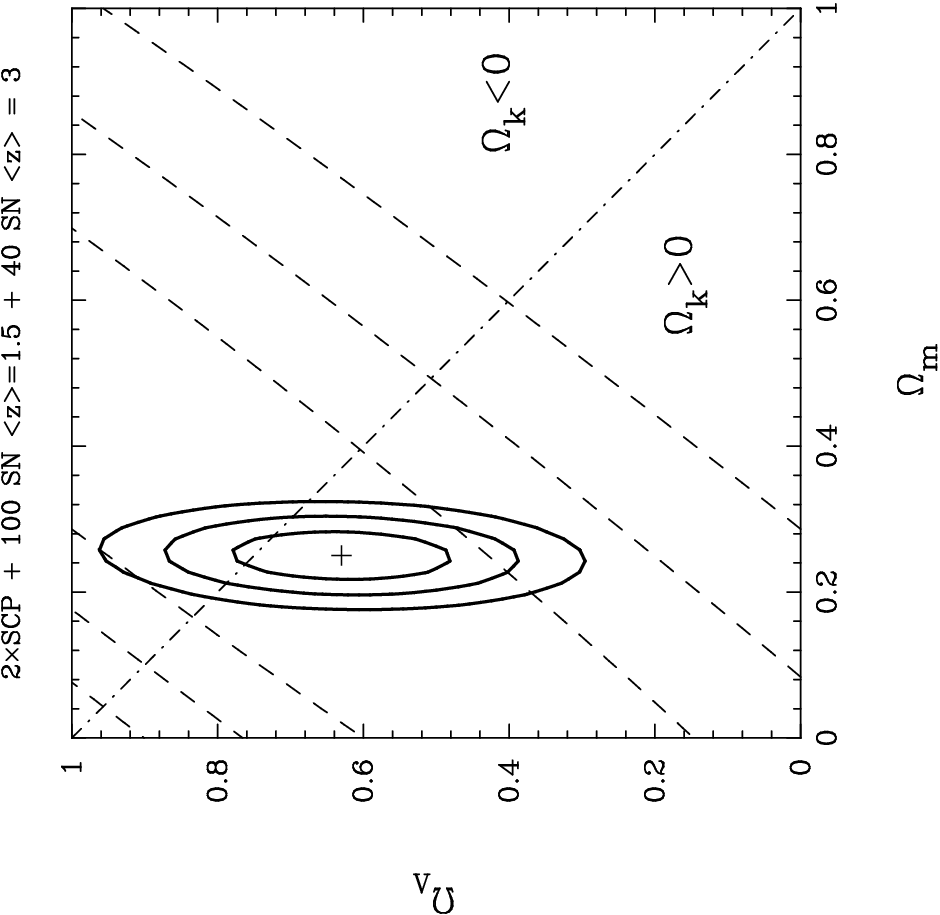}

\caption
{Fisher matrix constraints for a sample of SN extending to redshifts
$z>3$ (see text) illustrating that by extending the redshift range one
can determine $\Omega_m$ independently of $\Omega_\Lambda$.}

\label{figure3}
\end{figure}

\subsection{Constraining $\Omega_m$ with NGST}

Observations of very distant supernovae at $z \simgt 3$ may be
possible with a Next Generation Space Telescope ({\it e.g.}
Miralda-Escude and Rees 1997, Madau 1998, Livio 1999). We will not
analyse the feasibility of such observations here. Rather, we note
from Figures 1 and 2 that the major axis of the error ellipses in the
$\Omega_\Lambda$-- $\Omega_m$ tilt and become more vertical as the
redshift range of the SN sample is increased. This is because the
magnitude redshift relations for models with very different values
values of $\Omega_\Lambda$ and the same $\Omega_m$ converge at higher
redshifts. The convergence redshift depends on $\Omega_m$ and lies
between $z \approx 2$--$4$ for $\Omega_m$ in the range $0.2$--$1$ (see
Figure 1 of Melnick, Terlevich and Terlevich, 1999).

This is illustrated by Figure 3, which shows the $1$, $2$ and $3
\sigma$ likelihood contours determined from the Fisher matrix for a
sample consisting of twice the SCP sample, $100$ SN with $\langle z
\rangle = 1.5$, $\Delta z = 0.5$, and $40$ SN with $\langle z \rangle
= 3$, $\Delta z = 1$. As expected, these contours are almost vertical
in the $\Omega_m$--$\Omega_\Lambda$ plane. A sample of supernovae (or
some other distance indicator such as HII galaxies, Melnick \etal
1999) at redshifts $z \sim 3$ can therefore produce a tight constraint
on $\Omega_m$ independently of the value of
$\Omega_\Lambda$.

\section{Constraints on an Arbitrary Equation of State}
 
In this Section, we analyse the constraints that SN can place on an
arbitrary equation of state. We first consider a constant equation of
state. Models of this type (see Bucher and Spergel) include a
frustrated network of cosmic strings ($p/\rho = -1/3$) and a
frustrated network of domain walls ($p/\rho = -2/3$). A constant
equation of state is also a good approximation to a Q component
obeying tracker solutions. Tracker solutions are discussed in Section
3.2. Constraints on generalised forms of dark matter with
anisotropic stress are discussed by Hu \etal (1999) and will not be
considered here.

\subsection{Constant equation of state}

\begin{figure*}

\vskip 3.0 truein

\includegraphics{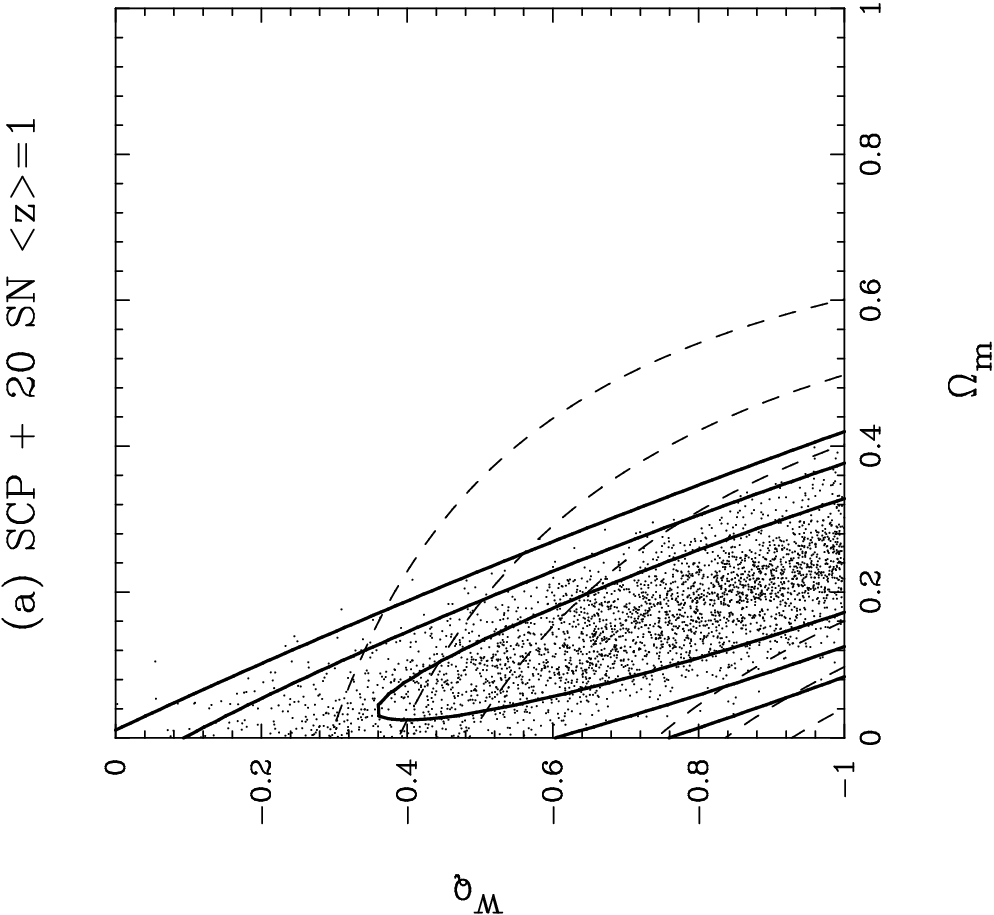}
\includegraphics{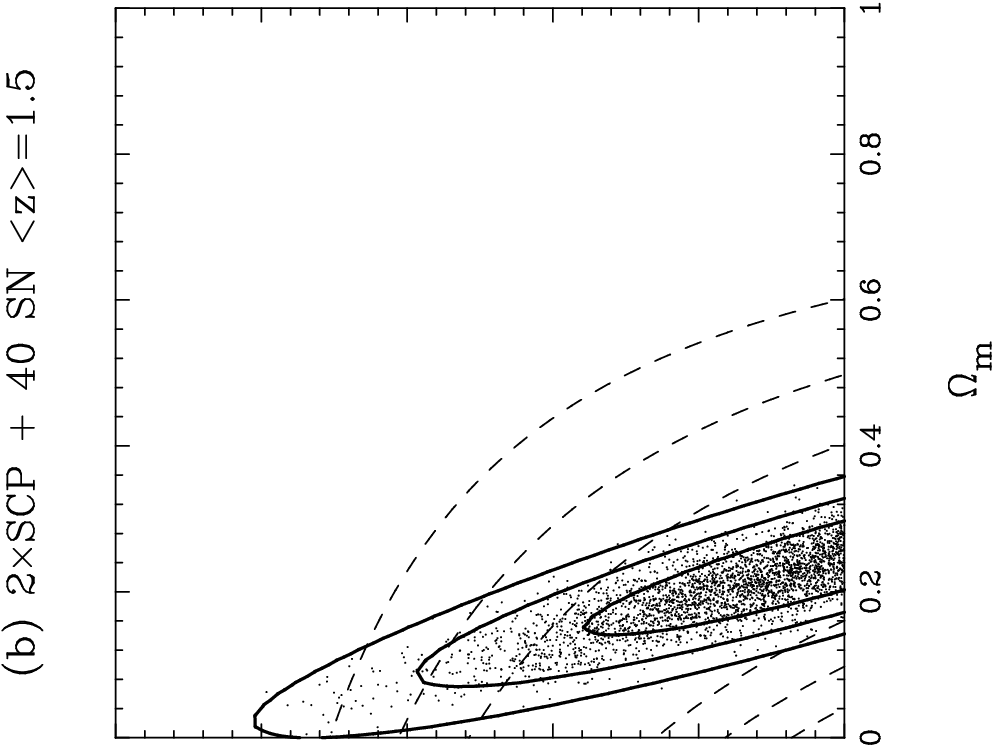}

\caption
{As Figure 2, but for an arbitrary constant equation of state in a
spatially
flat Universe.
The dashed lines in each panel show $1$, $2$ and $3 \sigma$ likelihood
contours in the $w_Q$--$\Omega_m$ plane for the SCP distant
supernova sample as analysed in Section 4 (assuming a constant
equation of state). The solid contours are derived from
the Fisher matrix for enhanced samples of high redshift supernovae
and the points show maximum likelihood derived from 
Monte-Carlo realizations of these samples.}
\label{figure4}
\end{figure*}

If we include a Q-like component with equation of 
state $p/\rho = w_Q$, the expression for the term $x$ in the 
luminosity distance (equation 2) is modified to
\begin{eqnarray}
x(z, \Omega_m, \Omega_Q, w_Q)  =  \qquad \qquad \qquad \qquad \qquad
\qquad \qquad \nonumber\\
 \int_0^z {dz^\prime 
\over 
[\Omega_m (1 + z^\prime)^3 + \Omega_k ( 1 + z^\prime)^2 +
\Omega_Q(1+z^\prime)^{3(1+w_Q)}]^{1/2} }. 
\end{eqnarray}

The addition of the parameter $w_Q$ means that it is not possible to
constrain all of the parameters $\Omega_m$, $\Omega_Q$, $w_Q$ to high
accuracy from the supernova data alone (see Section 4.2). Thus,
Garnavich \etal (1998) analyse the High-z Supernovae Search (HZS)
sample (Riess \etal 1999) assuming a spatially flat universe and find
that $w_Q < -0.55$ at $95\%$ confidence. A similar analysis of the
SCP sample by Perlmutter, Turner and White (1999) yields $w_Q \simlt -0.5$.

Table 2 lists the results of a Fisher matrix analysis for a Q-like
component with a constant $w_Q$. Here we have applied the constraints
$w_Q \ge -1$ and $\Omega_Q \ge 0$. The upper table gives results for
the supernovae magnitude-redshift relation alone assuming a spatially
flat Universe with $\Omega_m = 0.25$ and $w_Q = -1$. The constraints
on $w_Q$ from a sample such as the SCP data are quite poor and
improve relatively slowly as the sample is extended to higher
redshift  because of a strong degeneracy between
$w_Q$ and $\Omega_m$ in the magnitude-redshift relation. This is
illustrated in  Figure 4, which shows the analogue of Figure
2 for Q-like models. As the supernovae sample is extended to higher
redshift, the likelihood contours narrow but $w_Q$ and $\Omega_m$
remain strongly degenerate.

\begin{table}
\label{tab2}
\centerline{\bf Table 2: Fisher Matrix Errors, $\Omega_m$, $\Omega_Q$ and $w_Q$.}

\begin{center}
\begin{tabular}{|cccccc|} \hline
\multicolumn{6}{c} {Supernovae Alone, $\Omega_k = 0$} \\
              &  SCP   & \multicolumn{2}{c}{SCP + 20 SN} &
\multicolumn{2}{c}{2$\times$SCP + 40SN} \\
$<z>$                &          &  $1.0$ & 1.5 & 1.0 & 1.5 \\
$\delta \Omega_m$   &    $0.14\;\;$  & $0.12\;\;$ & $0.097$ & $0.097$ & $0.073$ \\ 
$\delta w_Q$        &    $0.36\;\;$  & $0.35\;\;$ & $0.32\;\;$ & $0.28\;\;$ & $0.24\;\;$\\
$\delta {\cal M}$   &    $0.051$ & $0.051$ & $0.051$ & $0.037$ & $0.036$\\\hline
\multicolumn{6}{c} {Supernovae +CMB, $\Omega_k = 0$} \\
              &  SCP   & \multicolumn{2}{c}{SCP + 20 SN} &
\multicolumn{2}{c}{2$\times$SCP + 40SN} \\
$<z>$                &          &  $1.0$ & 1.5 & 1.0 & 1.5 \\
$\delta \Omega_m$   &    $0.027$  & $0.022$ & $0.0210$ & $0.016$ & $0.015$ \\ 
$\delta w_Q$        &    $0.10\;\;$  & $0.085$ & $0.081$ & $0.061$ & $0.057$\\
$\delta {\cal M}$   &    $0.048$ & $0.046$ & $0.045$ & $0.032$ & $0.032$\\\hline
\multicolumn{6}{c} {Supernovae + CMB, $\Omega_k \ne 0$} \\
              &  SCP   & \multicolumn{2}{c}{SCP + 20 SN} &
\multicolumn{2}{c}{2$\times$SCP + 40SN} \\
$<z>$               &          &  $1.0$ & 1.5 & 1.0 & 1.5 \\
$\delta \Omega_m$   &    $0.14\;\;$  & $0.12\;\;$ & $0.095$ & $0.094$ & $0.069$ \\ 
$\delta \Omega_Q$   &    $0.10\;\;$  & $0.083$ & $0.066$ & $0.067$ & $0.048$ \\ 
$\delta w_Q$        &    $0.31\;\;$  & $0.31\;\;$ & $0.27\;\;$ & $0.24\;\;$ & $0.20\;\;$\\
$\delta {\cal M}$   &    $0.051$ & $0.051$ & $0.051$ & $0.036$ & $0.036$\\\hline
\end{tabular} 
\end{center} 
\end{table}

The situation is dramatically improved by the addition of constraints
from  CMB anisotropies. The addition of a Q-like component
affects the location of the Doppler peaks (see Caldwell \etal 1998,
White  1998) and, in analogy with equation (5), an accurate
determination of the CMB power spectrum imposes the constraint
\begin{equation}
\Delta \Omega_Q =  - {(\partial \gamma_D/\partial w_Q)
\over (\partial \gamma_D/\partial \Omega_Q)} \Delta w_Q -
{(\partial \gamma_D/\partial \Omega_m)
\over (\partial \gamma_D/\partial \Omega_Q)} \Delta \Omega_m,
\end{equation}
The second panel of Table 2 shows the constraints derived on an
arbitrary equation of state by combining supernovae data with the CMB 
constraint of equation (7). For spatially flat models, the
combination of SN and CMB anisotropies constrains  $w_Q$ to an
accuracy of better than $0.1$, sufficient to set tight constraints
on the physical parameters of  Q-like models (for example, whether 
one requires contrived potentials, see Section 4). However, the constraints 
on $w_Q$ improve relatively slowly as the SN sample is extended to
higher redshift. Similar conclusions apply if the assumption of a
spatially flat universe is relaxed (see the lower panel of Table 2). 
In that case, the parameters $\Omega_m$ and $\Omega_Q$ can be
determined to high precision, but the constraints on $w_Q$ improve
slowly as the SN sample is increased. This implies that it is worth
analysing the constraints on Q-like models with arbitrary spatial
curvature using current SN and CMB data (see Section 4.2).

\begin{figure}

\vskip 5.4 truein

\includegraphics{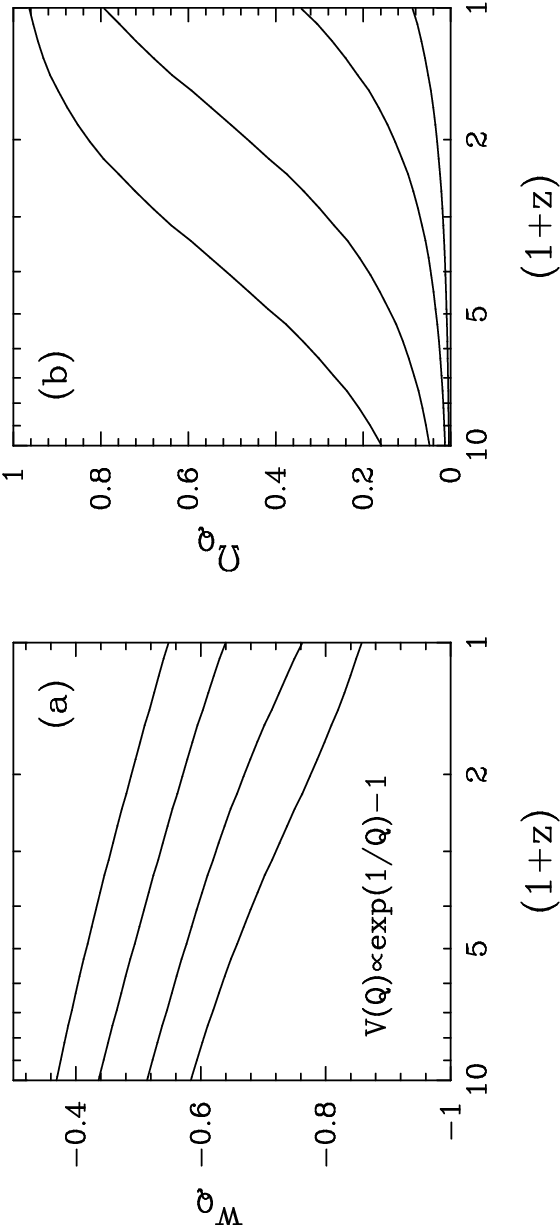}
\includegraphics{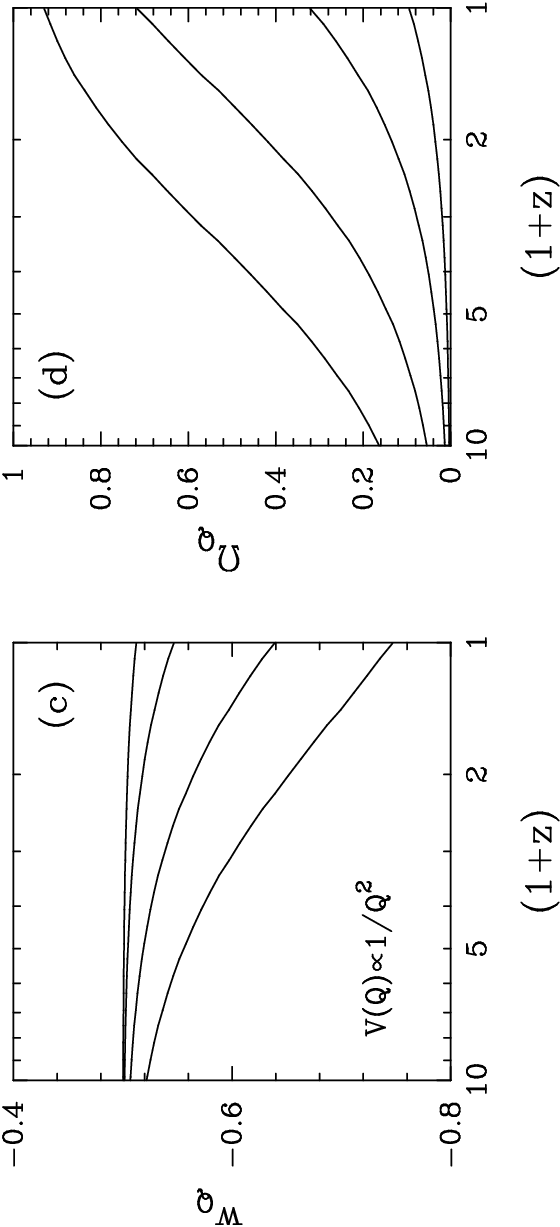}
\includegraphics{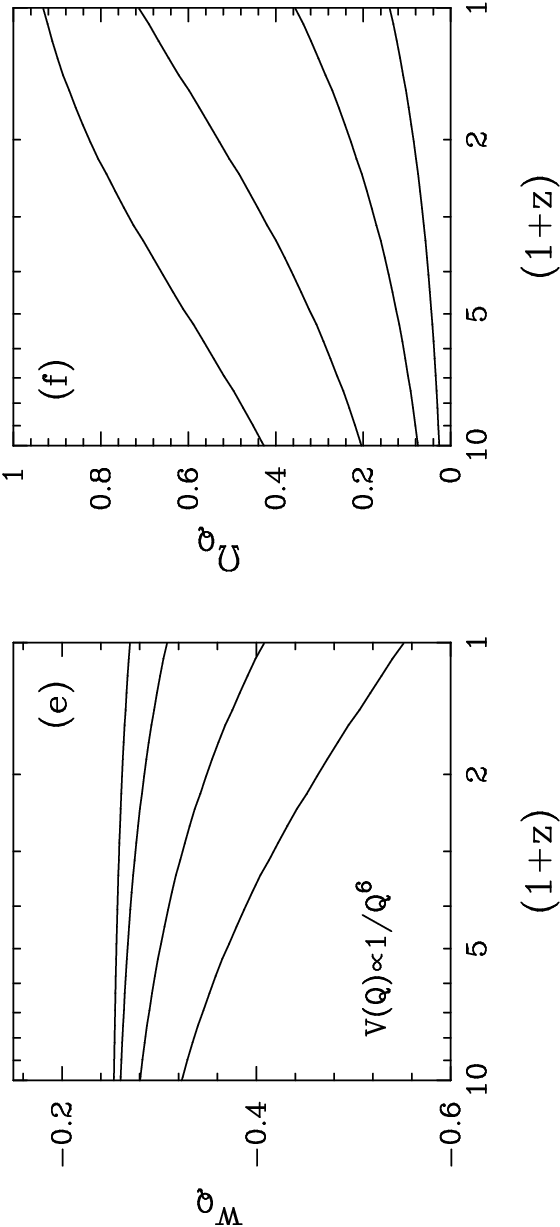}

\caption
{The evolution of the equation of state $w_Q$ and its contribution to
the cosmic density parameter $\Omega_Q$ as a function of redshift
derived from solutions to the tracker equation (8) for three
potentials: $V(Q) = M^4({\rm exp}(1/Q) - 1)$ (figures 5a and 5b); $V(Q)
= M^4(M/Q)^2$ (figures 5c and 5d); $V(Q) = M^4(M/Q)^6$ (figures 5e and
5f).  The curves in each figure are computed by varying the parameter
$M$, with more negative values of $w_Q$ corresponding to higher values
of $\Omega_Q$.}

\label{figure5}
\end{figure}

\begin{figure}

\vskip 5.4 truein

\includegraphics{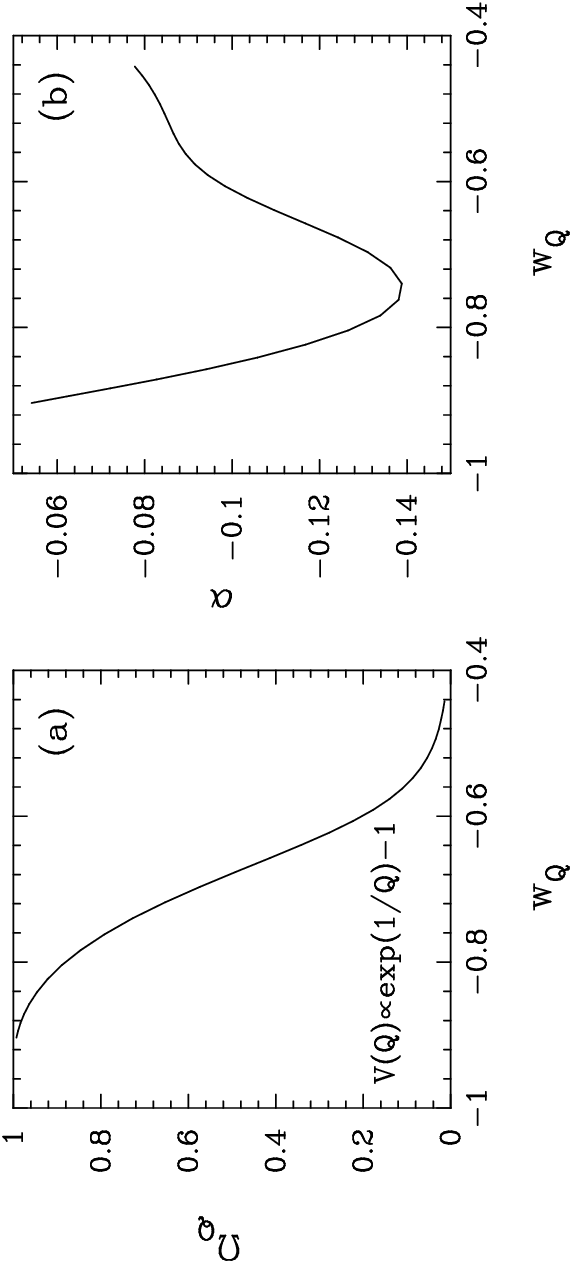}
\includegraphics{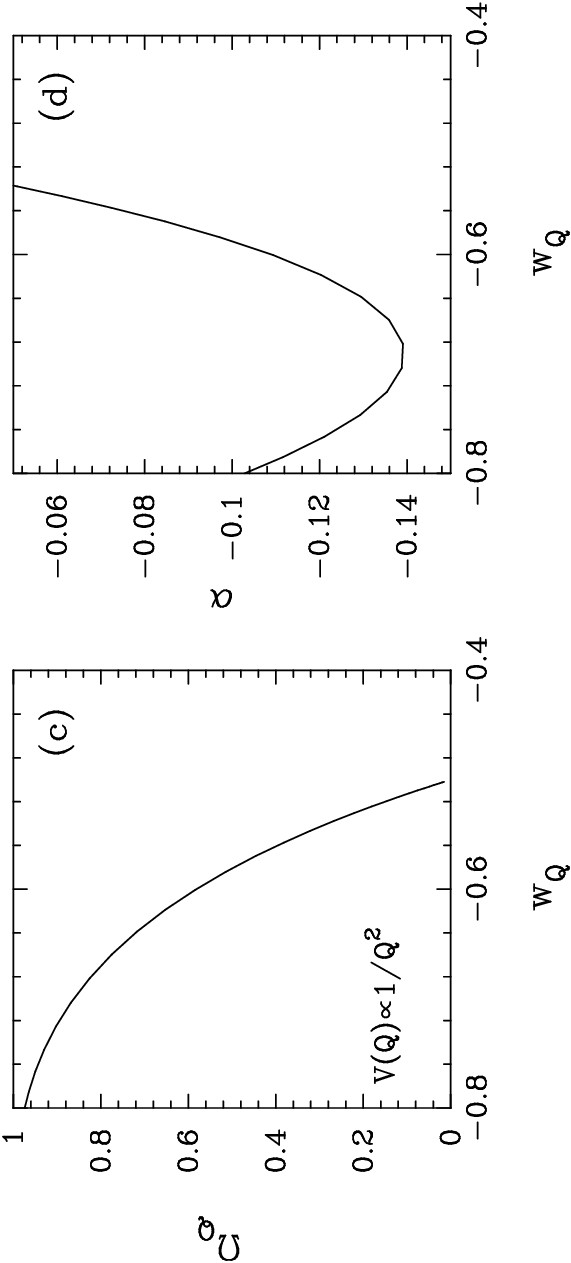}
\includegraphics{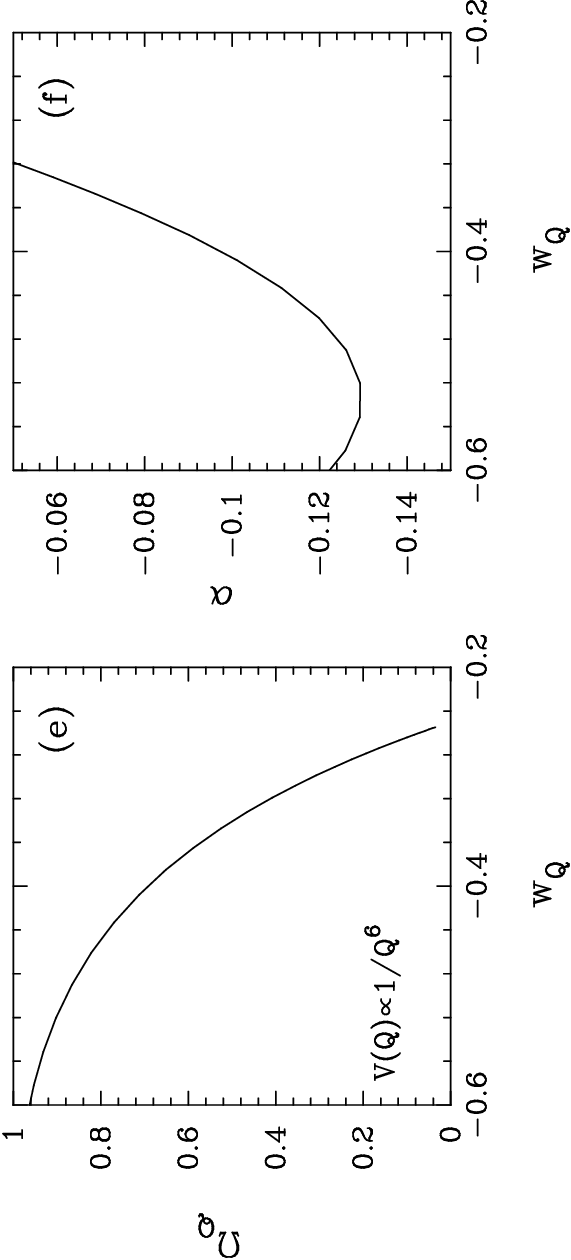}

\caption
{The left hand panels show the tracker solution relations between
$\Omega_Q$ and $w_Q$ at the present day for the three potentials used
in Figure 5. The right hand panels show the derivative $\alpha \equiv
\partial w_Q /\partial {\rm ln}a$ for the tracker solutions as a
function of $w_Q$.}

\label{figure6}
\end{figure}

\subsection{Time varying equation of state: tracker solutions}

In the previous section we have investigated the simplified case of a
constant $w_Q$. If, in fact, the Q-like component arises from a slowly
rolling scalar field evolving in a potential $V(Q)$, the equation of
state of the $Q$ component will vary  as a function of time. The
equations of motion of the $Q$ field can be written in the following
compact form (Steinhardt, Wang and Zlatev, 1998)
\begin{eqnarray}
{V^{\prime\prime} V \over (V^\prime)^2} =  1 + {w_B - w_Q \over 2 (1 + w_Q)}
- { 1 + w_B - 2w_Q \over 2 (1 + w_Q)} { \dot x  \over 6 + \dot x} \nonumber \\
 - {2 \over (1+w_Q)} {\ddot x \over ( 6 + \dot x)^2},  \qquad
x \equiv {(1+w_Q) \over (1 - w_Q)} \qquad 
\end{eqnarray}
where primes denote derivatives with respect to $Q$, $\dot x = d {\rm
ln}x/d {\rm ln a}$, $\ddot x = d^2 {\rm ln}x/d {\rm ln^2 a}$ and $a$
is the scale factor of the cosmological model. For a wide class of
potentials, and almost independently of the initial conditions, the
evolution of $Q$ locks on to a tracking solution in which $Q$ and
$w_Q$ vary slowly (see Zlatev, Wang and Steinhardt 1998, Steinhardt
\etal 1998). Examples of the evolution of $w_Q$ and $\Omega_Q$ at late
times are shown in Figure 5 for three forms of the potential
$V(Q)$. In each case, the evolution of $w_Q$ at $z \simlt 4$ is well
approximated by
\begin{equation}
w_Q = w_Q(a_0) + \alpha {\rm ln} (a/a_0)
\end{equation}
where $\alpha$ is a small number determined from the value of $\dot x$
at the present time.

Figure 7 shows the relations between $\Omega_Q$, $w_Q$ and $\alpha$ at
the present time derived from the solutions to equation 4
for the three potentials considered in Figure 5. The minimum value of
$\alpha$ is about $-0.14$, reflecting the fact that $Q$ is evolving
relatively slowly even at late times.

With the approximation of equation (9), the energy density of the $Q$
component evolves according to 
\begin{equation}
{\rho_Q(a) \over \rho_Q(a_0)} = \left ( {a \over a_0}
\right)^{-3(1+w_Q(a_0))} {\rm exp} \left( - {3 \over 2} \alpha [ { \rm
ln} (a/a_0) ]^2 \right).
\end{equation}
Note also that with the approximation of equation (9), the
tracker equation (8) becomes an algebraic equation relating
${V^{\prime\prime} V \over (V^\prime)^2}$ to $w_Q$, $\Omega_Q$ and
$\alpha$ ($w_B = w_Q \Omega_Q$ in the matter dominated era).

A small value of $\alpha \sim -0.1$ to $-0.2$ cannot be determined
accurately from SN and CMB observations because it is highly
degenerate with $w_Q$ and $\Omega_m$. As we will show in the next
Section, the introduction of the parameter $\alpha$ provides a
convenient way of testing the sensitivity of constraints on Q-like
models to the time evolution of $w_Q$.

\begin{figure*}

\vskip 3.0 truein

\includegraphics{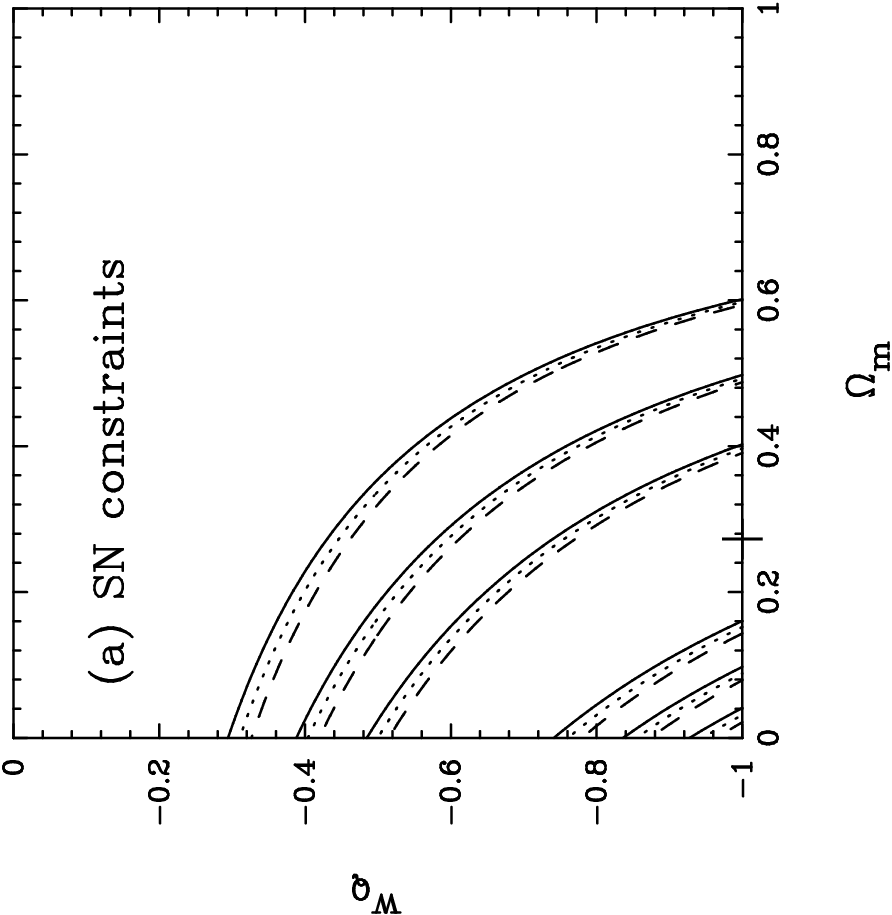}
\includegraphics{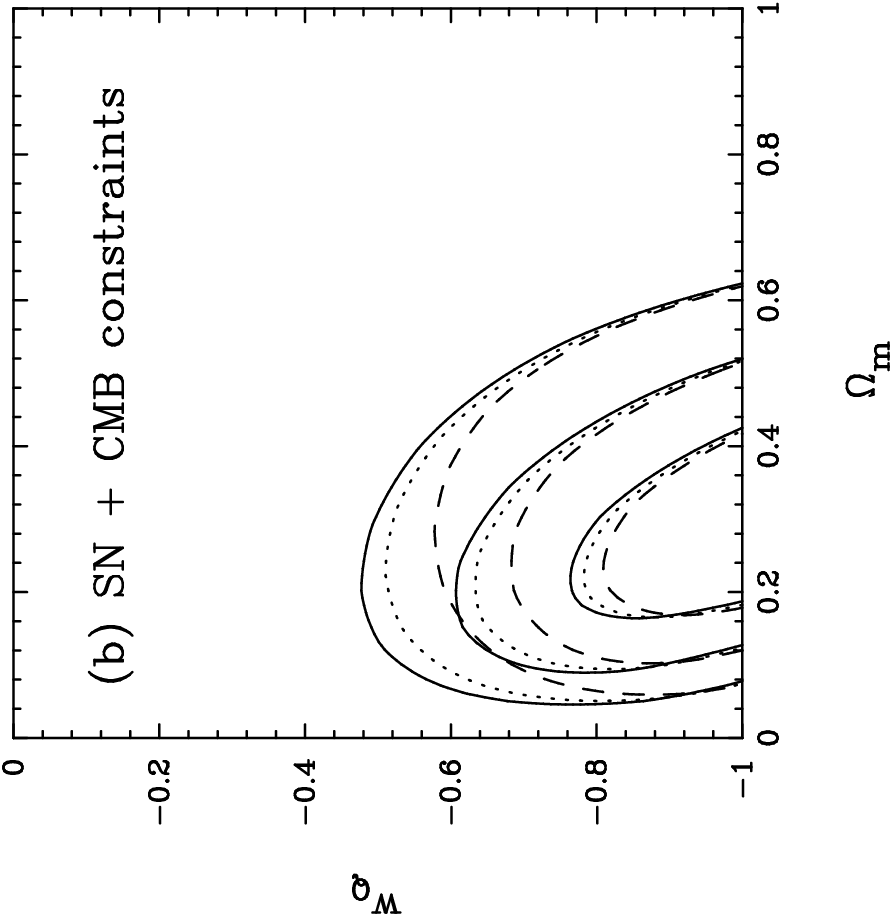}

\caption
{Constraints of $w_Q$ and $\Omega_m$ for spatially flat universes.
Figure 7a shows results for the SCP supernova sample following a
similar analysis to that presented by E99. Figure 7b shows results
for the supernova sample combined with the constraints from the CMB
anisotropy measurements as described in E99. The contours show $1$,
$2$ and $3 \sigma$ likelihood contours.The solid contours are derived
for $\alpha=0$, dotted contours are for $\alpha = -0.1$ and dashed
contours for $\alpha = -0.2$.}
\label{figure7}
\end{figure*}

We note that Huterer and Turner (1998) have recently proposed a
prescription for reconstructing the potential of a $Q$-like component
directly from the magnitude-redshift relation of Type Ia
supernovae. This approach may produce interesting constraints if the
field $Q$ is rapidly evolving at late times. For tracker solutions,
however, the equation of state changes so slowly that it would be
difficult to distinguish the true potential from a perfectly flat one.

\section{Limits on the equation of state 
from Type Ia Supernovae and the Cosmic Microwave Background}

\subsection{Spatially flat models}

In this Section, we use current SN and CMB data to constrain the
equation of state of the Universe. The analysis closely follows that
presented in E99. We use the sample of $56$ Type Ia SN of fit C of P98
and adopt the likelihood analysis described by E99 (including a
parametric fit to the luminosity-decline rate correlation), modifying the
expression for luminosity distance to incorporate the parameters of
the Q-like model. The CMB data that we use are plotted in Figure 1 of
E99.  We perform a likelihood analysis for these data assuming 
scalar adiabatic perturbations,  varying the amplitude of the
fluctuation spectrum, the scalar spectral index, the physical
densities of the CDM and baryons $\omega_c = \Omega_c h^2$, $\omega_b
= \Omega_b h^2$ \footnote{$h$ is the Hubble constant in units of $100\;
{\rm km} {\rm s}^{-1} {\rm Mpc}^{-1}$.},  and the Doppler peak location
parameter $\gamma_D$. Modifications to the CMB power spectrum arising
from spatial fluctuations in the Q component are ignored as these are
negligible in the slowly evolving $Q$ models considered here (see
Caldwell \etal 1998, Huey \etal 1998). We integrate over the CMB
likelihood assuming uniform prior distributions of the parameters to
compute a marginalized likelihood for $\gamma_D$ as described in E99.
The likelihood functions for the parameters $w_Q$, $\Omega_m$ and
$\Omega_Q$ presented below are constructed from the expression for the
angular diameter distance to the last scattering surface and the
probability distribution of $\gamma_D$.

Figure 7 shows the constraints on $w_Q$ and $\Omega_m$ for spatially
flat universes. The different line types show the constraints for
three different values of the parameter $\alpha$ characterising the
evolution of $w_Q$, $\alpha = 0$ (solid lines), $\alpha = -0.1$
(dotted lines) and $\alpha = -0.2$ (dashed lines).  As described in
the previous section, these values span the range found for tracker
solutions for a variety of potentials. These rates of evolution are so
low that they have very little effect on the likelihood contours. The
constraints plotted in Figure 7 are in very good agreement with those
derived by Garnavich \etal (1998) from an analysis of the HZS sample,
and with the analysis of the SCP sample (Perlmutter, Turner and White
1999) and of the combined HZS and SCP samples (Wang \etal 1999). The
fact that the constraints are weakly dependent on the size of the SN
sample is a consequence of the strong degeneracy between $w_Q$ and
$\Omega_m$ discussed in Section 3.2.

\begin{figure*}

\vskip 5.5 truein

\includegraphics{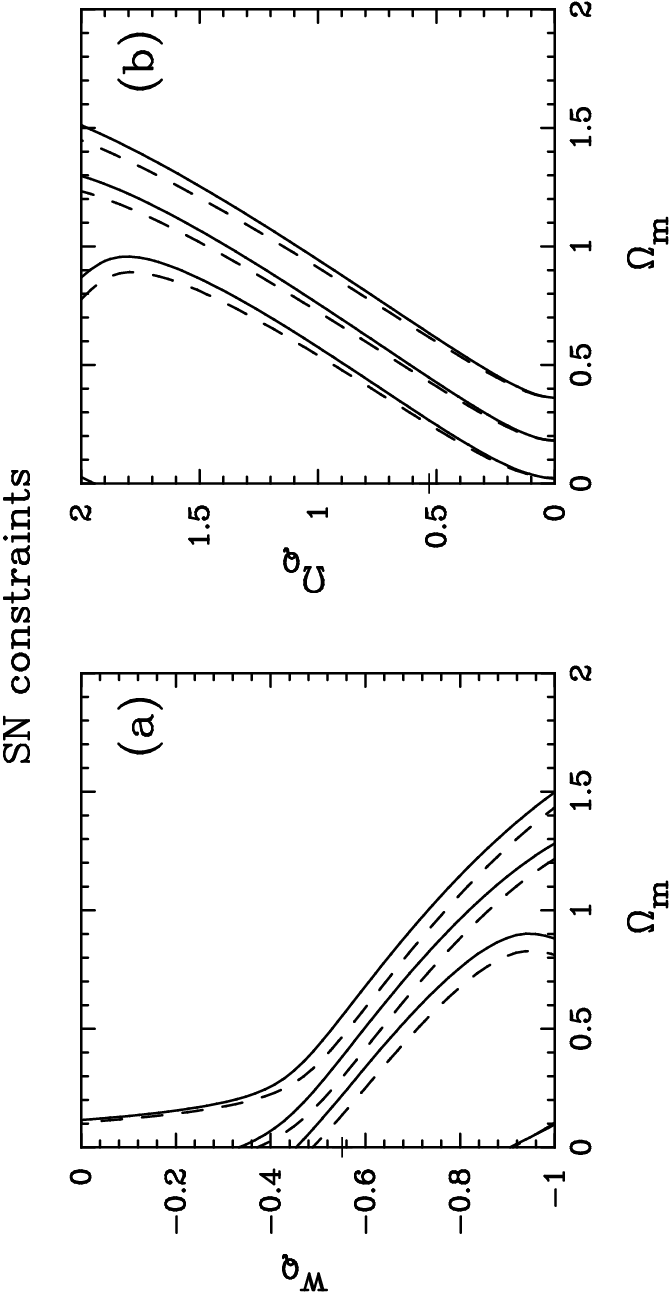}
\includegraphics{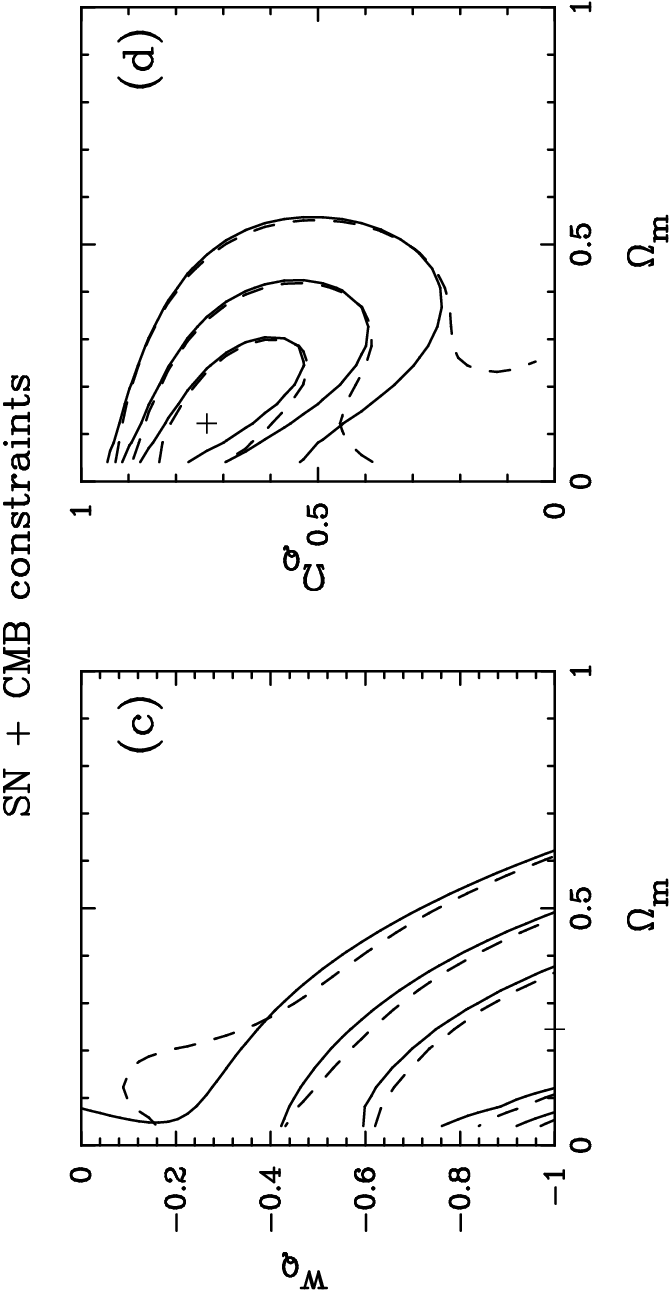}

\caption
{Analogue of Figure 5, but for quintessence models with arbitrary
spatial curvature. Figures 7a and 7b show marginalized likelihoods
in the $w_Q$--$\Omega_m$ and $\Omega_Q$--$\Omega_m$ planes derived
from Type Ia supernovae. Figures 7c and 7d show the combined
likelihoods for the Type Ia and CMB anisotropies. As in Figure 5,
the solid contours are derived for $\alpha=0$ and
dashed contours for $\alpha = -0.2$.}

\label{figure8}
\end{figure*}

Figure 7b shows the results of combining the SN likelihoods with those
determined from the CMB. The likelihood peaks at $w_Q = -1$, $\Omega_m
= 0.29$. Qualitatively, these results are similar to those of
Perlmutter \etal (1999); the favoured cosmology has an equation of
state $w_Q = -1$ and $w_Q$ is contrained to be less than $-0.6$ at the
$2 \sigma$ level. However, in detail, the constraints in Figure 7b are
somewhat less stringent than those of Perlmutter {\it et al.},, allowing a
broader range in $\Omega_m$ ($0.15 \simlt \Omega_m \simlt 0.5$ at the
$2 \sigma$ level). This is because Perlmutter \etal include
constraints on the power spectrum of galaxy clustering based on the
data compiled by Peacock and Dodds (1994)\footnote{Perlmutter \etal
(1999) do not combine the SN and CMB likelihoods but analyse the SN
data assuming a spatially flat Universe.}. In our view this is
dangerous because it requires a specific assumption concerning the
distribution of galaxies relative to the mass. Qualitatively, for
nearly scale-invariant adiabatic models, galaxy clustering imposes a
constraint on the parameter combination $\Gamma = \Omega_m h$ of $0.2
\simlt \Gamma \simlt 0.3$,  if galaxies are assumed to trace the mass
fluctuations on large scales (Efstathiou, Bond and White 1992, Maddox,
Efstathiou and Sutherland 1996). Combined with measurements of the
Hubble constant (for which Perlmutter \etal adopt $h \approx 0.65 \pm
0.05$), galaxy clustering leads to a constraint of $0.25 \simlt
\Omega_m \simlt 0.5$, partly breaking the degeneracy between $w_Q$ and
$\Omega_m$. The combined SN and CMB analysis in Figure 7b provides
constraints which are nearly as tight, but are much less model
dependent.

The constraints of Figure 7b place strong limits on Q-like models. For
tracking solutions, the constraint $\;w_Q \simlt$ $-0.6$ excludes steep
potentials ({\it e.g.} $V(Q) \propto Q^{-\beta}$ with $\beta \simgt
2$) and the data clearly favour a standard cosmological term ($w_Q =
-1$).  These limits on $w_Q$ are very close to the lower limit
($w_Q \simgt -0.7$) allowed for `physically well motivated' tracker
solutions (Steinhardt, Wang and Zlatev, 1998, {\it i.e.} 
smooth potentials with simple functional forms). With a slight
improvement of the observations one may be forced to fine-tune the
shape of the potential to construct a viable quintessence model.

The constraints of Figure 7b are somewhat stronger than those of Wang
\etal (1999), who perform a `concordance analysis' of Q-like models
using a number of observational constraints including those from Type
Ia supernovae and CMB anisotropies. These authors conclude limits of
$-1 \simlt w_Q \simlt -0.4$. The difference is caused by the different
methods of statistical analysis. The concordance analysis of Wang
\etal leads, by construction, to more conservative limits than the
maximum likelihood analysis and is more robust to systematic errors in
any particular data set. However, provided systematic errors are
negligible in the CMB and SN datasets, then the constraints of Figure
7b derived by combining likelihoods should be realistic.  These small
differences in the upper limits on $w_Q$ are important because they
can place significant restrictions on the physics.  As stressed in the
previous paragraph, the upper limit of $w_Q \approx -0.6$ places
strong constraints on tracker models with simple potentials.

\subsection{Models with arbitrary spatial curvature}

Figure 8 shows the results of a likelihood analysis of the SN and CMB
data, but now allowing arbitrary spatial curvature. We show two
projections of the likelihood distributions, marginalizing over
$\Omega_Q$ in Figures 8(a) and 8(c) and over $w_Q$ in Figures 8(b) and
8(c). The constraints, although weaker than those presented in Figure
7, are interesting nevertheless. The combined SN and CMB likelihoods
give a $2\sigma$ upper limit on $w_Q$ of $w_Q \simlt -0.4$ (Figure 8d)
and a maximum likelihood solution of $\Omega_m = 0.12$, $\Omega_Q =
0.73$ irrespective of the value of $w_Q$. Evidently, the SN and CMB
data constrain us to a nearly spatially flat Universe dominated either
by a cosmological constant, or a Q-like component with an equation of
state $w_Q \simlt -0.4$.

\section{Conclusions}

Observations of distant Type Ia supernovae have provided important
evidence that the Universe may be dominated by a cosmological constant
(P98, Riess \etal 1999). However, the constraints in the
$\Omega_\Lambda$--$\Omega_m$ plane from current data are degenerate
along a line defined by $\Omega_\Lambda \approx 0.32 + 1.43 \Omega_m$
(Figure 1). This degeneracy can be reduced significantly by extending
the redshift range of the supernovae sample. For example, with $20$
additional supernovae at redshift $z \sim 1.5$ the errors in
$\Omega_m$ and $\Omega_\Lambda$ could be reduced to $\delta \Omega_m
\approx 0.08$ and $\delta \Omega_\Lambda \approx 0.22$. A sample of
supernovae at $z \simgt 3$ could provide an accurate estimate of
$\Omega_m$ that is independent of the value of $\Omega_\Lambda$.

The combination of supernovae and CMB anisotropy measurements can
break the degeneracy between $\Omega_\Lambda$ and $\Omega_m$ if the
initial fluctuations are assumed to be adiabatic and characterised by
a smooth fluctuation spectrum. This method applied to recent
supernovae and CMB data suggests a nearly spatially flat universe
dominated by a cosmological term with $\Omega_\Lambda \approx 0.65$
(Lineweaver 1998, Garnavich \etal 1998, Tegmark 1999, E99). The only
plausible way of avoiding this conclusion is to appeal to some
systematic effect in the supernovae data, for example, grey dust or an
evolutionary effect in the supernovae data such as a metallicity
dependence (see {\it e.g.}  P98 for a discussion). The degeneracy
breaking afforded by extending the supernovae data to higher redshift
would provide an important consistency check of such systematic
effects and also on the interpretation of the CMB anisotropy data

The constraints on quintessence-like models with an equation of state
$w_Q = p/\rho$ improve relatively slowly as the supernovae data are
extended to higher redshift. The most promising way of constraining
$w_Q$ seems to be to combine supernovae and CMB measurements. We have
carried out a joint likelihood analysis of CMB anisotropy observations
and the SCP supernovae data. For a spatially flat Universe we derive a
$2\sigma$ upper limit of $w_Q = -0.6$. For universes of arbitrary
spatial curvature, the $2 \sigma$ upper limit is $w_Q = -0.4$.  The
combined SN and CMB likelihood peaks at $\Omega_m = 0.12$ and
$\Omega_Q = 0.73$ irrespective of the value of $w_Q$, suggesting that
the Universe is almost spatially flat.  The $2 \sigma$ upper limit of
$w_Q = -0.6$ for spatially flat Universes is close to the minimum value
of $w_Q \approx -0.7$ allowed for simple quintessence-models. This
suggests that some fine tuning of the potential may be required to
construct a viable quintessence model.

\vskip 0.2 truein

\noindent
{\bf Acknowledgements.} I thank Richard Ellis, Paul
Steinhardt and Roberto Terlevich 
for useful discussions and PPARC for the award of a Senior
Fellowship. I also thank Sarah Bridle, Anthony Lasenby, Mike Hobson
and Graca Rocha for allowing me to use their compilation of CMB
anisotropy data.

\end{document}